\documentclass[12pt]{article}

\usepackage[top=1in,bottom=1in,left=1in,right=1in]{geometry}

\usepackage{amsthm,amsmath}
\usepackage[utf8]{inputenc} % unicode support

\usepackage{graphicx}
\usepackage{helvet}
\usepackage{multirow}
\usepackage{bbm}
\usepackage{units}
\usepackage{url}
\usepackage{authblk}
\usepackage{color}
\usepackage{float}

\title{\textbf{The sequencing and interpretation of the genome obtained from a Serbian individual}}

\author[ ]{Wazim Mohammed Ismail$^1$\footnote{Contributed equally to this work.}, Kymberleigh A. Pagel$^{1*}$, Vikas Pejaver$^1$, Simo V. Zhang$^{1}$, Sofia Casasa$^2$, Matthew Mort$^3$, David N. Cooper$^3$, Matthew W. Hahn$^{1, 2}$, and Predrag Radivojac$^1$}

\affil[ ]{\footnotesize{$^1$Department of Computer Science, Indiana University, Bloomington, Indiana, U.S.A.}}
\affil[ ]{\footnotesize{$^2$Department of Biology, Indiana University, Bloomington, Indiana, U.S.A.}} 
\affil[ ]{\footnotesize{$^3$Institute of Medical Genetics, Cardiff University, Cardiff, U.K.}}

\date{\vspace{-5ex}} 

\begin{document}
\date{}
\maketitle

\begin{abstract}
Recent genetic studies and whole-genome sequencing projects have greatly improved our understanding of human variation and clinically actionable genetic information. Smaller ethnic populations, however, remain underrepresented in both individual and large-scale sequencing efforts and hence present an opportunity to discover new variants of biomedical and demographic significance. This report describes the sequencing and analysis of a genome obtained from an individual of Serbian origin, introducing tens of thousands of previously unknown variants to the currently available pool. Ancestry analysis places this individual in close proximity of the Central and Eastern European populations; i.e., closest to Croatian, Bulgarian and Hungarian individuals and, in terms of other Europeans, furthest from Ashkenazi Jewish, Spanish, Sicilian, and Baltic individuals. Our analysis confirmed gene flow between Neanderthal and ancestral pan-European populations, with similar contributions to the Serbian genome as those observed in other European groups. Finally, to assess the burden of potentially disease-causing/clinically relevant variation in the sequenced genome, we utilized manually curated genotype-phenotype association databases and variant-effect predictors. We identified several variants that have previously been associated with severe early-onset disease that is not evident in the proband, as well as variants that could yet prove to be clinically relevant to the proband over the next decades. The presence of numerous private and low-frequency variants along with the observed and predicted disease-causing mutations in this genome exemplify some of the global challenges of genome interpretation, especially in the context of understudied ethnic groups.
\end{abstract}

\section{Introduction}
The genetic variation between individuals accounts for much of observed human diversity and has the potential to provide information on phenotypic outcomes of clinical consequence. Studies of genetic variation provided by individual genome sequences have revealed that this variation differs both within and between populations, and also varies considerably depending upon the population \cite{Genomes2015}. Moreover, characterization of genetic variation of individuals from multiple populations has revealed a correlation between genetic and geographic distances, and has become relevant for determining genetic ancestry and geographic origin \cite{Novembre2008, Lazaridis2014}. Therefore, the characterization of genetic variation has been of major interest for diverse research fields, including medical, biological and anthropological sciences \cite{Burchard2003, Novembre2008, Gibson2009, Lazaridis2014, Lazaridis2016, Manrai2016}.

Sequencing of the first human genomes revealed that most genetic variation is derived from single nucleotide variants (SNVs), although insertions and deletions (indels) account for the majority of the variant nucleotides \cite{Levy2007}. The increased accessibility of DNA sequencing has contributed to individual efforts from a range of distinct populations. To date, individual genomes from American \cite{Levy2007, Wheeler2008}, Han Chinese \cite{Wang2008}, Russian \cite{Chekanov2010}, African \cite{Schuster2010}, Japanese \cite{Fujimoto2010}, German \cite{Suk2011}, Gujarati Indian \cite{Kitzman2011}, Estonian \cite{Lilleoja2012}, Pakistani \cite{Azim2013} and Mongolian \cite{Bai2014} populations have been sequenced and analyzed, among many others \cite{Genomes2015}.

Larger-scale efforts to characterize human genetic variation have demonstrated that individuals from different populations carry particular combinations of rare and low frequency variants. The 1000 Genomes Project Consortium has estimated that 86\% of all variants are confined to a single continental group and that about 10\% of variants observed in a population are private to that population \cite{Genomes2015}. Population-specific variants have the potential to be of both functional and biomedical importance \cite{Burchard2003}. Furthermore, evidence of biologically meaningful population-specific variation \cite{Guda2015} emphasizes the need for ethnically relevant reference genomes, as has been performed, for example, for the Korean population \cite{Cho2016}. Although we are not claiming to have introduced a new reference genome here, it is nevertheless important to expand our sequencing efforts across diverse populations, particularly those that have not been previously studied \cite{Popejoy2016, Manrai2016}.

In this paper, we describe the sequencing of the first genome of an individual of Serbian origin, a member of a relatively small population in Central to Southeastern Europe. We identify tens of thousands of novel genetic variants in this individual, more than a hundred of which map to  protein-coding regions and several hundred of which reside in close proximity to gene coding regions. The extent of observed genetic variation allowed comparisons with extant European populations and reaffirms support for the hypothesis of close correspondence between genetic and geographic distances \cite{Novembre2008}. These results contribute to ongoing efforts to understand human genetic variation and its geographic distribution, as well as placing the Serbian genome within the context of the broader European population structure. Testing for Neanderthal introgression in the genome, we find evidence suggesting gene flow from Neanderthal to an ancestral pan-European genome, with the Serbian genome being placed within range of other European populations. After variant annotation, we assess the burden of potentially pathogenic variation present in this genome and identify variants of putative clinical and pharmacogenetic relevance. Finally, we draw conclusions pertaining to the phenotypic consequences and biomedical interpretation of individually sequenced genomes. 

\section{Materials and Methods}

\label{sec:methods}

\subsection*{Donor information}
The individual whose genome was sequenced and analyzed is of Serbian descent. The data, both derived and raw, are publicly available through the Personal Genome Project website \cite{Ball2012}, participant ID: hu3BDC4B.

\subsection*{Sample collection and DNA sequencing}
Two milliliters of saliva were self-collected by the donor and stored using the DNA Genotek Oragene DISCOVER (OGR-500) sample collection kit. Extraction of DNA from the sample and subsequent sequencing were performed at the BGI (Shenzhen, China) on an Illumina HiSeq 2000 sequencer, using standard protocols. To minimize the likelihood of systematic bias in sampling, two libraries were prepared with an insert size of 500 bp each, with paired-end reads of length 90 bp. Sequencing was then carried out in four lanes for each library to ensure at least 30-fold coverage. 

\subsection*{Read mapping and variant calling}
The paired end reads were aligned to the GRCh37 human reference genome with the Burrows-Wheeler Aligner (BWA-MEM) \cite{Li2009a} and Bowtie2 \cite{Langmead2012}. Picard tools \cite{PicardTools} were used for the removal of duplicates followed by the use of two variant callers. GATK \cite{McKenna2010} was applied for base quality score recalibration, indel realignment, genotyping and variant quality score recalibration according to GATK Best Practices recommendations \cite{Auwera2013, DePristo2011}. We also generated results from Platypus \cite{Rimmer2014} applied to the output from each aligner. 

As described later in Results, variants identified through a combination of BWA and GATK were used for all analyses. Variants in the intersection of all four pipelines were considered to be confidently identified. All variant calls were subsequently annotated with information from NCBI RefSeq using ANNOVAR \cite{Wang2010}. 

\subsection*{Principal component analysis} 
Principal component analysis (PCA) was carried out using the $\mathsf{smartpca}$ program from EIGENSOFT (v6.0.1; https://github.com/DReichLab/EIG), on the Serbian genome combined with the SNV data (600,841 loci) from Lazaridis et al.~\cite{Lazaridis2014}. Only the subset of European individuals from their curated fully public dataset was used, reducing the original set of 1,964 individuals to 260. A projection to the first two principal components was used to establish the correspondence between genetic and geographic distance in our results.

\subsection*{Neanderthal introgression}
\label{sec:method_neanderthal}
To test for Neanderthal introgression in the Serbian genome, we computed D-statistics \cite{Green2010, Slatkin2016} using this genome and the dataset from Lazaridis et al.~\cite{Lazaridis2016}. This dataset includes 294 ancient individuals (only one of which was used here) and a diverse set of 2,068 present-day humans, genotyped on the Affymetrix Human Origins array. Both the archaic and modern genotype data were provided in the PACKEDANCESTRYMAP format, and were combined using the $\mathsf{mergeit}$ program from EIGENSOFT (v6.1.2; https://github.com/DReichLab/EIG). The merged dataset, in total, contains 2,362 samples genotyped at 621,799 SNV loci. As requested, we completed the consent form and obtained approval from David Reich's laboratory before using this dataset. We further note that some individuals from the study of Lazaridis et al.~\cite{Lazaridis2016} could not be included due to consent issues relating to data distribution. 

We next genotyped the Serbian genome against these predefined SNVs using GATK UnifiedGenotyper (v3.3-0-g37228af; \cite{DePristo2011}), following the same reference-bias free genotyping protocol as used by the Simons Genome Diversity Project \cite{Mallick2016}. We converted the resulting VCF files to the EIGENSTRAT format using VCFtools (v0.1.12a, \cite{Danecek2011}), and integrated the Serbian genotype with the modern and ancient datasets. Finally, we ran qpDstat from AdmixTools (default setting, v701) to calculate  D-statistics and to test for Neanderthal gene flow in the Serbian genome \cite{Slatkin2016}.

\subsection*{Burden of pathogenic variation}
Variants of putative clinical significance were identified using genotype-phenotype databases as well as computational variant-effect prediction. Manually curated genotype-phenotype databases, such as the Human Gene Mutation Database (HGMD) \cite{Stenson2017} and PharmGKB \cite{Klein2001}, annotate variants with a known relationship to phenotype \cite{Whirl2012, Stenson2017}. Clinical Annotations from PharmGKB were compared against dbSNP v142 rsIDs \cite{Sherry2001} obtained using the annotate\_variation.pl script in ANNOVAR and $\mathsf{avsnp142}$. Variants identified by GATK were compared against HGMD to identify potentially disease-causing and disease-associated mutations. 

All variants in protein-coding regions were extracted and inputted to the MutPred suite of tools \cite{Pejaver2017, Pagel2017, Mort2014}. The remaining variation observed in the proband was interrogated using CADD \cite{Kircher2014}. For disease and gene ontology associations, the hypergeometric test in WebGestalt was used with Benjamini-Hochberg correction for multiple hypothesis-testing \cite{Wang2013}. The background set that was used for these analyses included all protein-coding genes from the human reference genome. For significance of an ontology term, it was also required that at least five genes were associated with it.  
\section{Results}

\subsection*{Effect of genotyping software}
\label{sec:platform}
The selection of computational tools and their parameters in processing raw sequencing reads can significantly impact the resulting genome and the entirety of subsequent analysis. To understand the uncertainty of variant identification, we evaluated two different read mappers, BWA \cite{Li2009a} and Bowtie2 \cite{Langmead2012}, and two different variant callers, GATK \cite{McKenna2010} and Platypus \cite{Rimmer2014}. 

\begin{figure}[t]
\centering
\includegraphics[width=\linewidth]{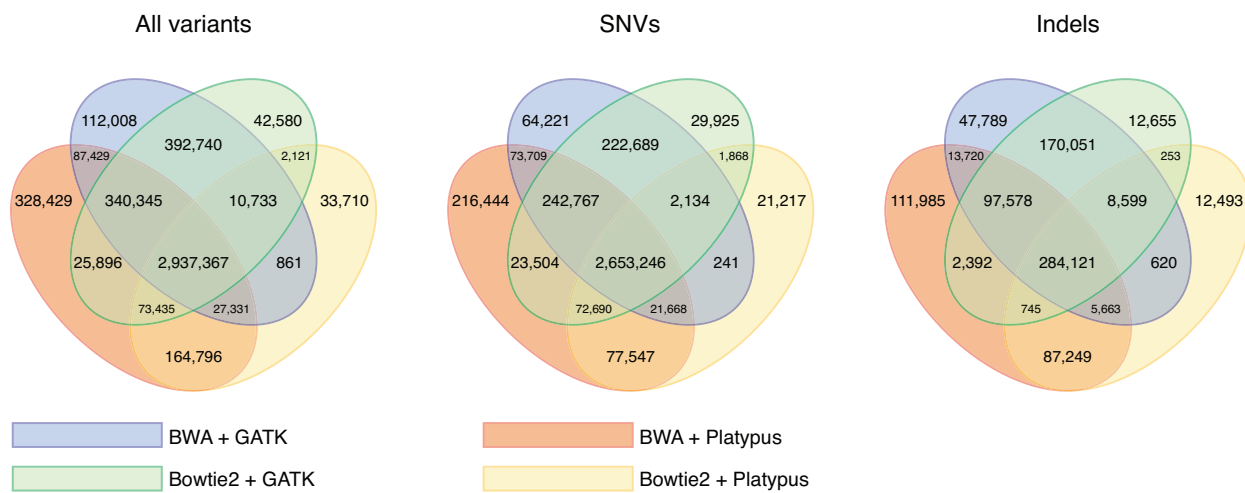}
\caption{Venn diagrams showing the total numbers of identified variants using two read mappers (BWA \cite{Li2009a}, Bowtie2 \cite{Langmead2012}) and two variant callers (GATK \cite{McKenna2010}, Platypus \cite{Rimmer2014}). }
\label{venn}
\end{figure}

The results from four different platforms are contrasted in Figure \ref{venn}. The SNV calling shows good concordance between both read mappers and variant callers, with a large proportion of variants identified by either platform being identified by all platforms. Using the BWA mapper, for example, 2,991,390/3,280,434 = 91.2\% of SNVs identified by GATK were also identified by Platypus and 89.1\% of SNVs identified by Platypus were also identified by GATK (Figure \ref{venn}). Indel calling, on the other hand, is less reliable, with 401,082/627,519 = 63.9\% variants identified by GATK also identified by Platypus and only 66.7\% of variants identified by Platypus being also identified by GATK. The influence of read mappers was markedly lower; i.e., using the GATK mapper, we found that 95.1\% of SNVs and 89.3\% of indels identified with BWA were also identified with Bowtie2, and that 98.3\% SNVs and of 97.6\% of indels identified with Bowtie2 were also identified with BWA. Smaller percentages of overlap were observed for Platypus. Based on the results observed in this work and the extent of usage of these tools in resequencing human genomes, we selected BWA$+$GATK as our main platform.

\subsection*{Identification of genetic variants}
The genome of a Serbian individual was sequenced according to the protocols described in Materials and Methods, %Section \ref{sec:methods}, 
with all 22 autosomes having similar coverage (and the X and Y chromosome having approximately half this coverage). The genome sequencing and mapping achieved an average read depth of 34.7, with 98.3\% of GRCh37 reference bases having coverage of 10-fold or more and 89.4\% having coverage of 20-fold or more. When only the non-zero positions were considered, the average read depth was 34.8, with 98.5\% of GRCh37 reference bases having coverage of 10-fold or more and 89.7\% having coverage of 20-fold or more. The number of zero-depth positions were 7,649,443 (0.3\%). The coverage distribution is shown in Supplementary Figure 1. 

Using the BWA$+$GATK pipeline, we identified a total of 3,908,814 variants in the Serbian genome, of which 2,195,638 (56.2\%) were heterozygous with one non-reference allele, 23,095 (0.6\%) were heterozygous with two non-reference alleles, and 1,690,081 (43.2\%) were homozygous for a non-reference allele. The reported variants passed all quality filters of GATK (marked as ``PASS'') and were subsequently mapped to GRCh37 human reference genomic regions using ANNOVAR \cite{Wang2010}. It is important to mention that ANNOVAR considers all heterozygous positions with both alternative alleles as two different variants. Therefore, the resulting genome contains 3,931,909 total variants, of which 2,940,042 (74.8\%) were identified by all four platforms and are considered to be confident identifications. Unsurprisingly, the majority of identified variants were found to reside in intergenic and intronic regions (Table~\ref{tab:breakdown}).

To identify novel variation, we searched the identified variants against the Genome Aggregation Database (gnomAD) \cite{Lek2016}. We found that 1.5\% (60,153) all variants and 0.4\% (12,439) of confident variants were not present in gnomAD. We shall refer to these variants as ``novel'' and ``confident novel'' variants, respectively. The breakdown of all variants and novel variants with respect to genomic location is shown in Tables~\ref{tab:breakdown}-\ref{tab:exonic}. The percentage of novel variants varied across categories, comprising 0.9\% (80) of nonsynonymous variants, 0.4\% of synonymous variants, 0.7\% (145) of exonic variants, 1.5\% (20,531) of intronic variants, and 1.6\% (34,779) of intergenic variants. We found that 45.0\% (9,328/20,739) of the exonic variants were nonsynonymous, whereas 50.1\% (10,381/20,739) were synonymous. Similar fractions were observed for the confident variants (44.1\% vs.~52.3\%). Of the 3,871,756 GATK variants which are also observed in the gnomAD database, 3,805,264 (95\%) of these variants are annotated to have allele frequency greater than 1\% in gnomAD and 3,676,638 (95\%) with allele frequency greater than 5\%.

\begin{table}[]
\centering
\caption{Summary of identified variants using BWA$+$GATK. Variants not present in gnomAD \cite{Lek2016} are listed as novel and variants identified by all four genotyping platforms are listed as confident.} 
\vspace{3mm}
\begin{tabular}{|l|l|l|p{2.2cm}|p{2.2cm}|}
\hline
\textbf{Type of Variant}        & \textbf{Variant} & \textbf{Novel}  & \textbf{Confident variants} & \textbf{Confident novel}  \\ \hline
upstream    & 23094  & 320  & 16211  & 90   \\ \hline
upstream; downstream      & 881    & 8    & 624    & 4    \\ \hline
UTR5        & 5205   & 54   & 4055   & 22   \\ \hline
UTR5; UTR3   & 16  & 1    & 12  & 0    \\ \hline
exonic      & 20706  & 145  & 17114  & 115  \\ \hline
exonic; splicing   & 33  & 1    & 22  & 0    \\ \hline
splicing    & 151    & 0    & 107    & 0    \\ \hline
intronic    & 1410507   & 20531   & 1078226 & 4336        \\ \hline 
UTR3        & 31066  & 409  & 24095  & 101  \\ \hline
downstream  & 26685  & 398  & 19351  & 61   \\ \hline 
ncRNA\_exonic  & 13064  & 129  & 9520   & 30   \\ \hline
ncRNA\_exonic; splicing   & 3   & 0    & 2   & 0    \\ \hline
ncRNA\_intronic   & 235936    & 3376    & 173168        & 832  \\ \hline
ncRNA\_splicing   & 65  & 1    & 51  & 0    \\ \hline
ncRNA\_UTR5    & 1   & 1    & 0   & 0    \\ \hline
intergenic  & 2164496   & 34779   & 1597484       & 6848        \\ \hline
\end{tabular}
\label{tab:breakdown}
\end{table}

\begin{table}[]
\caption{Summary of identified exonic variants using BWA+GATK. Variants not present in gnomAD \cite{Lek2016} are listed as novel and variants identified by all four platforms are listed as confident.} 
\vspace{3mm}
\centering
\begin{tabular}{|l|l|l|p{2.2cm}|p{2.2cm}|}
\hline
\textbf{Type of Variant} & \textbf{Variants} & \textbf{Novel} 
& \textbf{Confident variants} & \textbf{Confident novel}\\ \hline
synonymous SNV           & 10381 & 42 & 8965 & 36 \\  \hline
nonsynonymous SNV        & 9328& 80 & 7559 & 69 \\ \hline
nonframeshift deletion   & 137& 2 & 62 & 0 \\ \hline
nonframeshift insertion  & 117& 3 & 58 & 0 \\ \hline
frameshift deletion      & 103& 6 & 45 & 4 \\ \hline
frameshift insertion     & 74 & 3 & 37 & 1 \\ \hline
stopgain                 & 87 & 6 & 54 & 4 \\ \hline
stoploss                 & 11 & 0 & 9 & 0 \\ \hline
unknown                  & 501 & 4 & 347 & 1  \\ \hline
\end{tabular}
\label{tab:exonic}
\end{table}

\subsection*{Genetic variation and geographic distance}

\begin{figure}[t]
\centering
\includegraphics[width=6.5in]{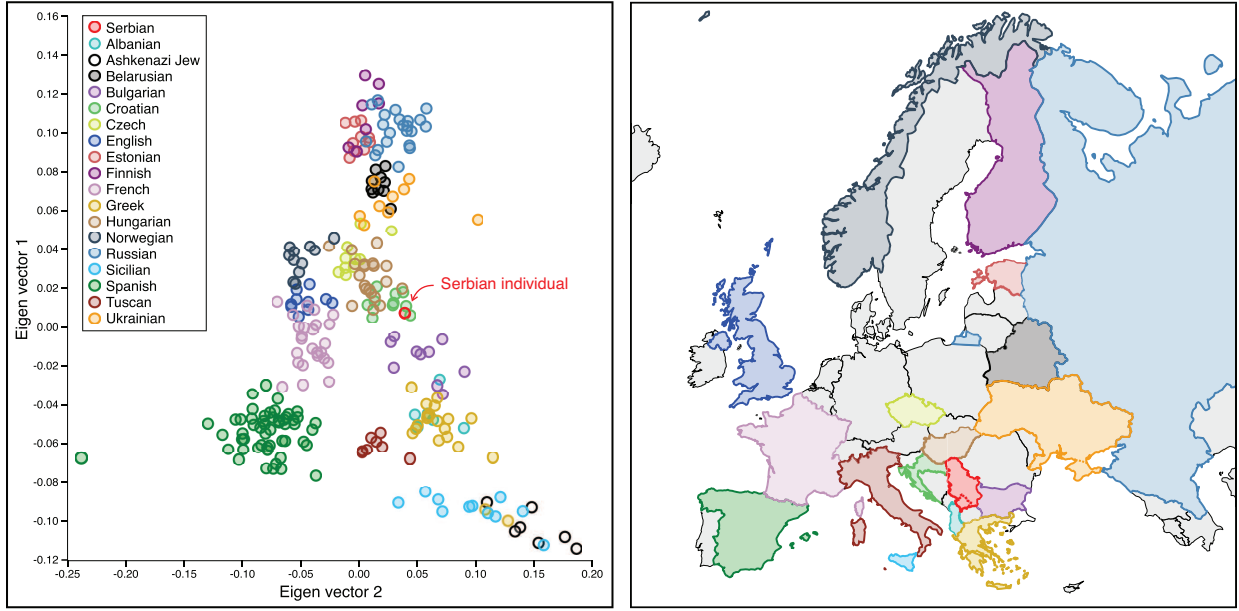}
\caption{Principal component analysis (PCA) plot showing the proximity of the genome sequenced in this study to other European genomes. As observed in previous studies \cite{Novembre2008, Lazaridis2014}, genomic distance correlates with geographic distance.}
\label{ancestry}
\end{figure}

The projection of the Serbian individual to the first and second principal components against European groups from \cite{Lazaridis2014} confirms that individuals from the same geographic region cluster together (Figure \ref{ancestry}). We clearly distinguish clusters of major populations composed of individuals from the same region, approximately mirroring a map of Europe. The PCA plot demonstrates that the genetic ancestry of the Serbian individual analyzed in the present study corresponds to its geographic distance from other populations. It is positioned in close proximity of the Croatian, Bulgarian, and Hungarian populations. 

A somewhat surprising finding is the similarity of distances between the Serbian individual and other mostly Slavic populations (Russian, Belarus, Ukrainian) relative to distances to various Central, Western, and Southern European groups (Czech, French, English, Albanian, Greek). The average Euclidean distance and variance between the Serbian individual and each of the available populations in the two-dimensional space of major PCA components is as follows: Croatian ($0.016826 \pm 0.010526$), Bulgarian ($0.033603 \pm 0.000225$), Hungarian ($0.037121 \pm 0.000177$), Czech ($0.053687 \pm 0.000033$), Albanian ($0.058875 \pm 0.000117$), Ukrainian ($0.064328 \pm 0.000062$), Belarusian ($0.069803 \pm 0.000043$), Greek ($0.071108 \pm 0.00062$), Tuscan ($0.0736441 \pm 0.000028$), French ($0.083077 \pm 0.000159$), English ($0.084570 \pm 0.000142$), Norwegian ($0.092721 \pm 0.00088$), Russian ($0.095968 \pm 0.000079$), Estonian ($0.098421 \pm 0.000046$), Finnish ($0.108523 \pm 0.000154$), Sicilian ($0.120370 \pm 0.000481$), Spanish ($0.134602 \pm 0.000776$), Ashkenazi ($0.156692 \pm 0.000538$). The three closest individuals to the Serbian genome were of Croatian ancestry (0.0038, 0.0046, and 0.0108).

We note that combining the Serbian individual with the set of 260 European individuals from Lazaridis et al.~\cite{Lazaridis2014} caused 50 formerly biallelic sites to become triallelic (no monoallelic sites became triallelic). The triallelic sites were removed from the analysis, leaving 600,791 sites in the analysis. The $\mathsf{smartpca}$ program was applied to the 261-by-600,791 genotype matrix.

%Figure 1

\subsection*{Gene flow with Neanderthals}
Comparisons between Neanderthals and modern humans have previously revealed evidence of gene flow from Neanderthals to Europeans \cite{Green2010, Slatkin2016, Prufer2014, Durand2011}. To test whether the Serbian genome shares an excess of alleles with the Neanderthal genome, we integrated the Serbian genotype with a published panel of ancient and modern humans (Materials and Methods). %(Section \ref{sec:method_neanderthal}). 
We calculated  D-statistics as a formal test for gene flow based on a four-taxon phylogeny, $D(P_1, P_2, P_3, O)$, where $P_i$ ($i \in \{1,2,3\}$) are populations and $O$ is an outgroup. Given a scenario where gene flow is absent, the derived alleles of $P_3$ are expected, with equal likelihood, to match those of $P_1$ and $P_2$; i.e., $D=0$. Alternatively, either $P_1$ or $P_2$ could share alleles with $P_3$ more often than not, in which case $D$ deviates from zero.

We computed $D(\textrm{Yoruba}, \textrm{Serbian}, \textrm{Altai}, \textrm{Chimpanzee})$ for testing for gene flow between Neanderthals (``Altai'') and the given Serbian genome. We expected a positive $D$ value, given previous evidence that Neanderthals exchanged more  alleles with Europeans than with Africans. The test returned a $D$ value of $0.0241 \pm 0.004476$, which significantly deviated from zero (Z-score~=~5.39; Table~\ref{tab:neanderthal}), suggesting gene flow between Neanderthal and the lineage leading to the Serbian genome. To validate this result, we also ran the test for other European populations (Table~\ref{tab:neanderthal}). D-statistics calculated for French, Greek, and Russian genomes were comparable to our result, all falling within the expected range of values reported in previous studies \cite{Green2010, Prufer2014, Durand2011}.

We further attempted to ensure that the calculated D-statistics were unbiased. To do so, we repeated the analysis by replacing Yoruba with Mbuti, as some of the Yoruba samples could have had some recent European admixture. The calculation for $D$(Mbuti, Serbian, Altai, Chimpanzee) yielded a $D$ value of $0.0186 \pm 0.004763$ (Z-score~=~3.99; Table~\ref{tab:neanderthal}), consistent with our results using the Yoruba samples. We next checked whether the Serbian individual has reference biases in genotyping that could have inflated the $D$ value. We performed D-statistics tests in the form of $D$(other European population, Serbian, Mbuti, hg19ref) and chose French, Greek and Russian as the ``other European population''. We obtained no test results indicating the bias of Serbian genotypes toward the reference (French: $0.0038 \pm 0.004078$; Greek: $0.0090 \pm 0.004182$; Russian: $0.0074 \pm 0.004192$).

\begin{table}
\footnotesize
\caption{Testing gene flow with Neanderthals. The results show the D-statistic ($D$), its standard error (SE) and Z-score ($Z$) for the test using the set of populations $P_1$, $P_2$, and $P_3$, with Chimpanzee as an outgroup ($O$). The last two columns show ABBA vs.~BABA counts over the four genomes ($P_1$, $P_2$, $P_3$, $O$).}
\vspace{3mm}
\begin{centering}
\begin{tabular}{|c|c|c|c|c|c|c|c|c|}
\hline 
$P_{1}$ & $P_{2}$ & $P_{3}$ & $O$ & $D$ & SE & Z-score & ABBA & BABA\tabularnewline
\hline 
\hline 
Yoruba & Serbian & Altai & Chimpanzee & 0.0241 & 0.004476 & 5.393 & 18158 & 17302\tabularnewline
\hline 
Yoruba & French & Altai & Chimpanzee & 0.0266 & 0.003012 & 8.821 & 18284 & 17338\tabularnewline
\hline 
Yoruba & Greek & Altai & Chimpanzee & 0.0270 & 0.003034 & 8.906 & 18266 & 17305\tabularnewline
\hline 
Yoruba & Russian & Altai & Chimpanzee & 0.0288 & 0.003096 & 9.306 & 18328 & 17302\tabularnewline
\hline 
Mbuti & Serbian & Altai & Chimpanzee & 0.0186 & 0.004763 & 3.909 & 18817 & 18129\tabularnewline
\hline 
Mbuti & French & Altai & Chimpanzee & 0.0210 & 0.003532 & 5.941 & 18902 & 18125\tabularnewline
\hline 
Mbuti & Greek & Altai & Chimpanzee & 0.0214 & 0.003578 & 5.978 & 18897 & 18106\tabularnewline
\hline 
Mbuti & Russian & Altai & Chimpanzee & 0.0232 & 0.003600 & 6.434 & 18932 & 18074\tabularnewline
\hline 
\end{tabular}
\par\end{centering}
\label{tab:neanderthal}
\end{table}

\subsection*{Analysis of medically relevant variants}
The sequenced genome contains 2,344 genetic variants that are present in HGMD by virtue of their having been previously associated with a risk of disease; the proportions of variants within each effect category are shown in Table~\ref{HGMD_summary}. Several homozygous variants, manually annotated as disease-causing (DM) are observed in the genome, shown in Table~\ref{HGMD_DM}. Of these, two are youth-onset phenotypes that are homozygous for the disease-causing allele in the proband genome: nephropathic cystinosis (NP\_004928.2:p.T260I) and Factor XIII deficiency (NM\_000129.3:c.-19+12C$>$A). The disease phenotypes associated with these homozygous mutations typically become apparent in childhood, and therefore their occurrence in a healthy adult is indicative of variable penetrance. The other homozygous disease-causing variants result in phenotypes that have not yet been observed in either the individual or in that individual's family history, perhaps reflecting either low expressivity or late-onset. Observed heterozygous disease-causing mutations are primarily childhood-onset without presentation in the individual, although they may represent recessive conditions; thus, their failure to manifest may not necessarily be indicative of poor reporting or curation quality.

\begin{table}[]
\centering
\footnotesize
\caption{Amount of disease-causing and potentially disease-relevant variation in the Serbian genome. Identified variants were searched against HGMD and broken down into the phenotypic categories of HGMD. Variants were broken down into exonic and noncoding as well as homozygous and heterozygous.}
\vspace{3mm}
\begin{tabular}{|p{8cm}|c|c|c|c|}
\hline 
 & \multicolumn{2}{c|}{\textbf{Exome}} & \multicolumn{2}{c|}{\textbf{Noncoding}} \\ \hline
& \multicolumn{1}{l|}{\textbf{Hom}} & \multicolumn{1}{l|}{\textbf{Het}} & \multicolumn{1}{l|}{\textbf{Hom}} & \multicolumn{1}{l|}{\textbf{Het}} \\ \hline
Disease-causing mutations (DM)  & 2 & 9 & 4 & 6 \\ \hline
Likely disease-causing mutations (DM?) & 29 & 51 & 8 & 31 \\ \hline
Disease-associated polymorphisms with additional supporting functional evidence (DFP) & 78 & 139 & 203 & 301 \\ \hline
Disease-associated polymorphisms (DP)  & 233 & 356 & 189 & 322 \\ \hline
Polymorphisms that affect gene/protein structure, function or expression but with no reported disease association (FP)  & 63 & 95 & 95 & 130 \\ \hline
\end{tabular}
\\
\begin{flushleft} {\footnotesize The number of homozygous and heterozygous variants that are associated with variants reported in HGMD. HGMD labels correspond to the strength and/or evidence for the relationship between variant and disease. }
\end{flushleft}
\label{HGMD_summary}
\end{table}

We also identified several variants of potential pharmacogenetic relevance using PharmGKB. Variants in PharmGKB are assigned Clinical Annotation Levels of Evidence from variants with preliminary evidence  (Level 4) to high confidence variant-drug combinations with medically endorsed integration into health systems (Level A1). The genome contains a single variant with a high confidence annotation (Level 1B): rs2228001, associated with toxicity and adverse drug reaction to cisplatin, a chemotherapeutic agent. A further 17 variants were annotated with moderate evidence to impact the dosage, efficacy, metabolism and/or toxicity of drugs for diverse phenotypes including chronic hepatitis C, organ transplantation rejection, glaucoma, depression, schizophrenia, asthma, epilepsy and HIV infections, as well as several chemotherapy drugs. 

\begin{table}[]
\centering
\footnotesize
\caption{Disease-causing variants observed in the proband. The table contains analysis of six homozygous variants form the sequenced genome that are listed in HGMD as disease-causing.}
\vspace{3mm}
\begin{tabular}{|l|l|l|p{4.4cm}|} \hline 
\textbf{Gene} & \textbf{Variant} & \textbf{rsID} & \textbf{Phenotype} \\  \hline
\emph{MIR137HG} & NC\_000001.10:g.98502934G$>$T & rs1625579 & Schizophrenia increased risk  \\  \hline
\emph{SLC12A3} & NM\_000339.2:c.1670-8C$>$T & NA & Gitelman syndrome without hypomagnesaemia  \\  \hline
\emph{DUOXA2} & NM\_207581.3:c.554+6C$>$T & NA & Hypothyroidism  \\  \hline
\emph{F13A1} & NM\_000129.3:c.-19+12C$>$A & rs2815822	& Factor XIII deficiency  \\  \hline
\emph{CTNS} & NP\_004928.2:p.T260I & NA & Cystinosis nephropathic  \\  \hline
\emph{PNPLA2} & NP\_065109.1:p.P481L & rs1138693 & Myopathy late-onset  \\  \hline
\end{tabular}
\label{HGMD_DM}
\end{table}

\subsubsection*{Pathogenicity prediction} 
In addition to known disease-associated variants, we identified missense variants predicted to be pathogenic by MutPred2 \cite{Pejaver2017}. Of the 11,206 missense variants called by GATK, 9,329 passed all quality filters (annotated as `PASS'). Of these, 9,305 variants were unambiguously mapped to the correct protein isoforms and hence were amenable for prediction by MutPred2. Based on a score threshold of 0.8 (estimated 5\% false positive rate), 95 missense variants were predicted to be `pathogenic.' Of these, 14 variants were found in the homozygous state and 81 were found in the heterozygous state. Genes for these variants were enriched in GO terms related to peptidase activity (Supplementary Figure 2). A similar analysis for disease associations revealed that the subject may be at risk for cardiovascular disorders (Supplementary Table 1). 

We assessed the pathogenicity of 180 nonsense and frameshifting insertion and deletion variants with MutPred-LOF \cite{Pagel2017}. From this set, we identified a total of 7 variants with scores above the 0.5 score threshold (corresponding to a 5\% false positive rate) (Supplementary Table 2).  We also assessed the pathogenicity of the 90 SNV splicing variants with MutPred Splice \cite{Mort2014}. Of these, 28 of the variants are scored at least 0.6 and were therefore classified as a ``Splice Affecting Variant'' by MutPred Splice. Of these, one variant is predicted to cause loss of natural 3' splice sites, two variants are predicted to interrupt cryptic 3' splice sites, and three variants are predicted to disrupt cryptic 5' splice sites. These 28 variants are described in Supplementary Table 3. 

To ensure assessment of the complete variome of the proband, we utilized CADD v1.3 \cite{Kircher2014} to evaluate all noncoding variants. To do so, we utilized a scaled C-score cutoff of 20 to identify the 1\% most damaging variants. In total, we found 16 UTR variants, 1,630 intronic variants, 3,911 intergenic variants, 80 regulatory variants, 839/533 upstream/downstream variants, and 9 variants annotated as ``noncoding\_change.'' All of these were predicted to be deleterious. The noncoding variants with the highest C-scores are described in Supplementary Table 4.  
\section{Discussion}

This work describes the first whole-genome sequencing of a Serbian individual. Ancestry analysis positioned the Serbian individual in closest proximity to the Croatian population, consistent with its South Slavic ancestry \cite{Kushniarevich2015}. Our analyses further support the hypothesis of gene flow between Neanderthal and pan-European ancestral populations, with the level of introgression into the Serbian genome being within the range observed in other European populations. Previous genetic studies involving Slavic populations employed mitochondrial, Y-chromosome and SNV-panel data to investigate the relationship between geographic, genetic and linguistic distances \cite{Kushniarevich2015, Davidovic2017}. Consistent with this work, our analyses expand the scope beyond Slavic populations and further contribute to the understanding of human genetic variation and its geographic distribution.

In contrast to studies using genotyping arrays \cite{Novembre2008, Lazaridis2014, Kushniarevich2015, Davidovic2017}, the availability of whole-genome sequences presents the opportunity for a high-resolution individualized analysis. To this end, we found that the sequenced genome contains a significant number of previously unobserved variants, which emphasizes the importance of continued sequencing of a large number of individuals, especially from smaller ethnic groups. Subsequent sequencing of other Serbian individuals could provide further insight into these novel variants; e.g., whether they are private to the population or to the individual. Such results would in turn contribute important information regarding variants that are currently considered to be rare, with implications for improved variant interpretation. Furthermore, new algorithms and reduced sequencing costs will have the potential to provide higher-quality analysis of structural variants that has not been possible in this work. Our analysis also found a number of variants of clinical and pharmacogenomic significance that might extend beyond an individual's disease risks to facilitate possible future medical interventions. Such variants might contribute to better outcomes in studies of disease penetrance, mechanistic understanding of population risks, and database curation.

Recent advances in high-throughput sequencing and reduced costs of genotyping have greatly facilitated whole-genome data generation, and have become key to understanding both human phenotypes and early human history \cite{Novembre2008, Lazaridis2014}. However, modern technology and cost structure continue to pose challenges in determining and interpreting one's genome \cite{vanNimwegen2016}. Variation in read mapping and variant calling contribute to the uncertainty of interpretation with different software packages, identifying different sets of variants. We found that inter-software discrepancies ranged from relatively small for SNVs to considerable for insertions and deletions. Therefore, variant and genome interpretation demand caution, since thousands of SNVs and tens of thousands of indels may simply constitute genotyping errors \cite{Nielsen2011, Wall2014}.

It is worth mentioning that in addition to the technical aspects of genome sequencing, an important aspect of genome interpretation concerns psychosocial uncertainty due to phenotypic and privacy-associated risks \cite{Wang2017}. This study has provided demographic analysis that supports the individual's own sense of Serbian ancestry; however, the finding of multiple predicted youth-onset pathogenic mutations in a healthy individual provides cautionary lessons for predictive medicine.

\bibliographystyle{plain}
\bibliography{main}

\end{document}

\maketitle

\section{HGMD annotations}
There are five different classes of variant listed in HGMD. Disease-causing mutations (DM) are entered into HGMD where the authors of the corresponding report(s) have established that the reported mutation(s) are involved (or very likely to be involved) in conferring the associated clinical phenotype upon the individuals concerned. The DM classification may, however, also appear with a question mark (DM?), denoting a probable/possible pathological mutation, reported as likely to be disease causing in the corresponding report, but where (i) the author has indicated that there may be some degree of doubt or uncertainty; (ii) the HGMD curators believe greater interpretational caution is warranted, or (iii) subsequent evidence has appeared in the literature which has called the initial putatively deleterious nature of the variant into question (e.g. a negative functional, case–control or population-scale sequencing study). In addition, three categories of polymorphism are included in the database. Disease-associated polymorphisms (DP) are entered into HGMD where there is evidence for a significant association with a disease/clinical phenotype along with additional evidence that the polymorphism is itself likely to be of functional relevance (e.g., as a consequence of genic location, evolutionary conservation, transcription factor binding potential, etc.), although there may be no direct evidence (e.g., from an expression study) for a functional effect. The functional polymorphisms (FP) class includes those sequence changes for which a direct functional effect has been demonstrated (e.g., by means of an \emph{in vitro} reporter gene assay or alternatively by protein structure, function or expression studies), but with no disease association reported as yet. Disease-associated polymorphisms with supporting functional evidence (DFP) must meet both of the above criteria in that the polymorphism should not only have been reported to be significantly associated with disease, but should also display direct evidence of being of functional relevance.

\begin{figure}[H]
\centering
\includegraphics[angle=270,width=6.5in]{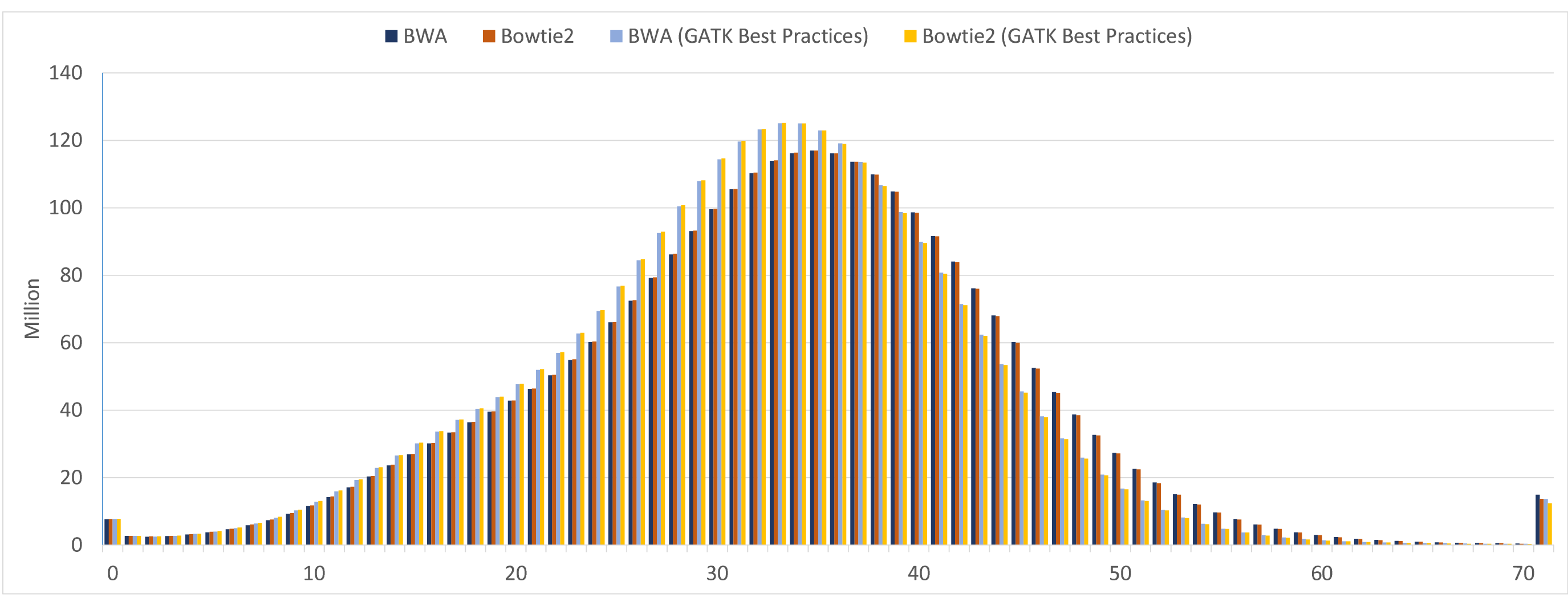}
\caption{Histogram of read depths. The last bar includes all read depths greater than 70.}
\label{depth}
\end{figure}

\begin{figure}[H]
\centering
\includegraphics[width=5.2in]{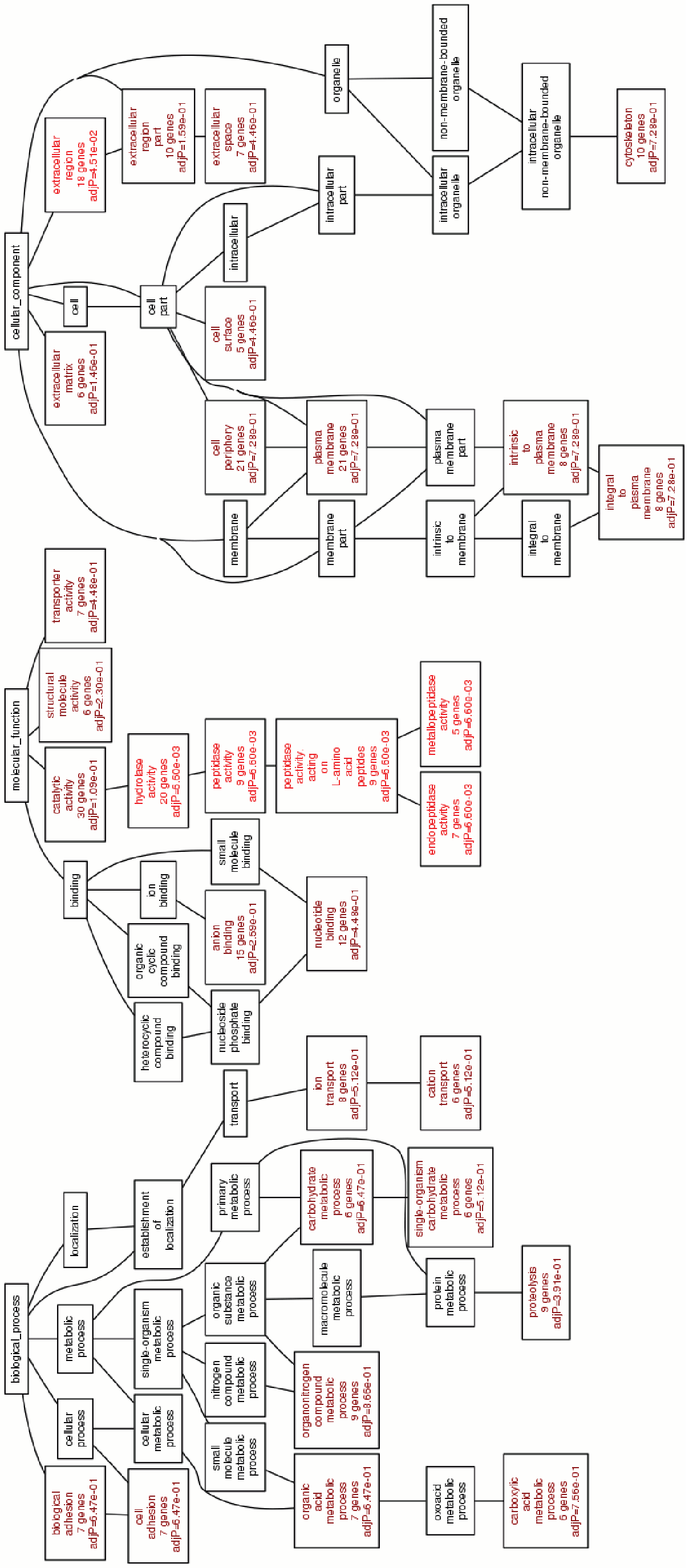}
\caption{GO terms enriched in the set of 81 genes that harbored the 95 missense variants predicted to be pathogenic (highlighted in red).}
\label{goterms}
\end{figure}

%Supplementary Table 3
\begin{table}[]
\centering
\footnotesize
\caption{Top 5 diseases enriched in the set of 83 genes containing predicted pathogenic missense variants}
\vspace{3mm}
\begin{tabular}{|l|p{1.5cm}|p{5.5cm}|p{2.8cm}|}
\hline
\textbf{Disease term} & \textbf{Number of genes} & \textbf{Entrez gene IDs} & \textbf{Benjamini-Hochberg adjusted $P$-value} \\  \hline
Renal diabetes & 5 & 55748, 1636, 2646, 9844, 6580 & 0.0005 \\ \hline
Cardiovascular Diseases & 8 & 4153, 350, 10060, 3816, 1636, 4624, 2646 2153 & 0.0007 \\ \hline
Ischemia & 5  & 4153, 10060, 9620, 1636, 2153 & 0.0007 & \hline
Respiratory Tract Diseases & 7 & 4153, 27294, 5270, 1285, 1636, 27159 2153 & 0.0007 & \hline
Lung Diseases  & 7  & 4153, 27294, 5270, 1285, 1636 & 0.0007 \\ \hline
\end{tabular}
\label{mp2_het}
\end{table}

\begin{table}[]
\centering
\caption{Frameshifting insertion/deletion variants scored with MutPred-LOF}
\vspace{3mm}
\footnotesize
\begin{tabular}{|l|l|l|p{9cm}|}
\hline
\textbf{Gene} & \textbf{Variant} & \textbf{Score} & \textbf{Affected Molecular Mechanisms} \\ \hline
NM\_030665 & Q280Hfs*84 &	0.57394	& Methylation(p=0.003); Amidation(p=0.004); Ubiquitylation(p=0.004); MoRF(p=0.005); O-linked glycosylation(p=0.007)  \\ \hline
NM\_030665 & Q280Afs*108 & 	0.57334	& Methylation(p=0.003); Amidation(p=0.004); Ubiquitylation(p=0.004); MoRF(p=0.005); O-linked glycosylation(p=0.007) \\ \hline
NM\_024675 & R170Ifs*14 & 	0.54289	& MoRF(p=0.03); Pyrrolidone carboxylic acid(p=0.033); Loop(p=0.034); Amidation(p=0.036); O-linked glycosylation(p=0.038) \\ \hline
NM\_006580 & A56Lfs*16 & 	0.52228	& Palmitoylation(p=0.02) \\ \hline
NM\_001037333 & Q95Pfs*15 &	0.511	& Helix(p=0.015); PPI hotspot(p=0.038) \\ \hline
NM\_001035235 & V110Dfs*2  &	0.50882	& NA  \\ \hline
NM\_001009931 &  M1Cfs*28 &	0.50536	& Signal cleavage(p=0); RNA binding(p=0); Sulfation(p=0); B factor(p=0.0001); Copper binding(p=0.0002) \\ \hline 
\end{tabular}
\label{mutpredlof}
\end{table}

\begin{table}[]
\centering
\caption{Splice variants scored with MutPred Splice}
\vspace{3mm}
\footnotesize
\begin{tabular}{|l|l|p{4cm}|l|p{4cm}|}
\hline
\textbf{Gene} & \textbf{Variant}  & \textbf{Genomic Coordinates} & \textbf{Score} & \textbf{Confident  Hypotheses }\\ \hline 
NP\_001091994.1  & I229T & chr19,17648350,+ & 0.68 &  \\ \hline
NP\_775815.2 & I229T & chr19,17648350,+ & 0.68 &  \\ \hline
NP\_115702.1 & Q65P & chr4,944210,+ & 0.78 &  \\ \hline
NP\_001727.1 & D2N & chr19,47823038,+ & 0.8 & Loss of natural 3' SS \\ \hline
NP\_702918.1 & T218T & chr9,131235321,+ & 0.81 &  \\ \hline
NP\_702915.1 & T218T & chr9,131235321,+ & 0.81 &  \\ \hline
NP\_002531.3 & T213T & chr9,131235321,+ & 0.81 &  \\ \hline
NP\_702914.1 & T237T & chr9,131235321,+ & 0.81 &  \\ \hline
NP\_702917.1 & T262T & chr9,131235321,+ & 0.81 &  \\ \hline
NP\_702913.1 & T301T & chr9,131235321,+ & 0.81 &  \\ \hline
NP\_001229281.1 & T232T & chr9,131235321,+ & 0.81 &  \\ \hline
NP\_001229283.1 & T156T & chr9,131235321,+ & 0.81 &  \\ \hline
NP\_702911.1 & T237T & chr9,131235321,+ & 0.81 &  \\ \hline
NP\_001229282.1 & T237T & chr9,131235321,+ & 0.81 &  \\ \hline
NP\_702910.1 & T281T & chr9,131235321,+ & 0.81 &  \\ \hline
NP\_000322.2 & T77S & chr11,18290880,+ & 0.83 &  \\ \hline
NP\_954630.1 & T77S & chr11,18290880,+ & 0.83 &  \\ \hline
NP\_001171477.1 & T77S & chr11,18290880,+ & 0.83 &  \\ \hline
NP\_065946.2 & A357A & chr19,35506729,+ & 0.86 & Cryptic 3' SS \\ \hline
NP\_001129671.1 & A350A & chr19,35506729,+ & 0.87 & Cryptic 3' SS  \\ \hline
NP\_001018101.1 & Q175Q & chr15,57918090,+ & 0.87 &  \\ \hline
NP\_001018100.1 & Q175Q & chr15,57918090,+ & 0.87 &  \\ \hline
NP\_689664.3 & Q175Q & chr15,57918090,+ & 0.87 &  \\ \hline
NP\_001018110.1 & Q175Q & chr15,57918090,+ & 0.87 &  \\ \hline
NP\_055146.1 & S481S & chr4,139100372,- & 0.88 &  \\ \hline
NP\_001185763.2 & A1757T & chr1,144868170,- & 0.91 & Cryptic 5' SS \\ \hline
NP\_055459.4 & A1757T & chr1,144868170,- & 0.91 & Cryptic 5' SS  \\ \hline
NP\_001185761.1 & A1651T & chr1,144868170,- & 0.91 & Cryptic 5' SS \\ \hline
\end{tabular}
\label{splice}
\end{table}

%\vspace{3mm}

\footnotesize
\begin{longtable}[]{|l|l|p{7.2cm}|l|l|}
\caption{Top scoring noncoding variants from CADD}
\hline 
\label{cadd}  %\begin{tabular}{|l|l|p{2.6cm}|p{2.6cm}|l|l|}'
\textbf{Chr} & \textbf{Pos}  & \textbf{Ref/Alt} & \textbf{Consequence}   & \textbf{Score} \\ \hline
17 & 59667953 & G/A & noncoding\_change & 37\\ \hline
16 & 88780634 & AGGTGTG/A & downstream & 35  \\ \hline 
2 & 85570766 & CAGAT/C & upstream & 35 \\ \hline
17 & 40270029 & ACTGAAGCTGAGGAGAGAGAGAGACGT CAGGGATGGGGGG/A & downstream & 27\\ \hline
8 & 145617534 & TGGGGGTGCAAGGTGA/T & downstream & 25.3\\ \hline
17 & 59668021 & G/C & noncoding\_change & 24.4\\ \hline
22 & 37465385 & CTGGGG/C & regulatory & 24.4\\ \hline
14 & 77648237 & TGTG/T & regulatory & 24.3\\ \hline
2 & 61389737 & T/G & noncoding\_change & 23.8\\ \hline
19 & 44610843 & C/T & noncoding\_change & 23.8\\ \hline
6 & 160560897 & CTGGTAAGT/C & regulatory & 23.5\\ \hline
2 & 20448521 & C/T & upstream & 23.5\\ \hline
3 & 10368577 & C/A & upstream & 23.3\\ \hline
12 & 104341644 & AACTG/A & downstream & 23.2\\ \hline
3 & 6513628 & CACAG/C & intergenic & 23.2\\ \hline
3 & 96336236 & G/T & upstream & 23.2\\ \hline
9 & 113128561 & TGAGGGTAG/T & downstream & 23.1\\ \hline
15 & 37172064 & C/A & downstream & 23.1\\ \hline
4 & 8785222 & C/G & intergenic & 23.1\\ \hline
7 & 11874349 & TA/T & upstream & 23.1\\ \hline
17 & 36782034 & TAAA/T & upstream & 23.1\\ \hline
21 & 35795163 & T/A & upstream & 23.1\\ \hline
4 & 24516813 & TTTATC/T & downstream & 23\\ \hline
7 & 131185995 & C/A & downstream & 23\\ \hline
10 & 103002063 & G/A & downstream & 23\\ \hline
11 & 61165731 & C/CA & downstream & 23\\ \hline
15 & 38982825 & C/CT & downstream & 23\\ \hline
15 & 100514418 & C/G & downstream & 23\\ \hline
17 & 5286861 & GTAGTGTTTGGAATTTTCTGTTCATA/G & downstream & 23\\ \hline
1 & 91621787 & TAACA/T & intergenic & 23\\ \hline
4 & 156503230 & ATATT/A & intergenic & 23\\ \hline
5 & 130484813 & T/A & intergenic & 23\\ \hline
6 & 14806096 & GTGACAT/G & intergenic & 23\\ \hline
6 & 85335098 & T/A & intergenic & 23\\ \hline
9 & 96708343 & A/G & intergenic & 23\\ \hline
11 & 36910028 & TAA/T & intergenic & 23\\ \hline
14 & 25819285 & CAT/C & intergenic & 23\\ \hline
21 & 16468682 & TACAA/T & intergenic & 23\\ \hline
3 & 64430551 & G/C & upstream & 23\\ \hline
6 & 32083175 & C/T & upstream & 23\\ \hline
10 & 62493816 & G/A & upstream & 23\\ \hline
11 & 93584075 & T/A & upstream & 23\\ \hline
12 & 114843606 & C/T & upstream & 23\\ \hline
16 & 87638576 & G/A & upstream & 23\\ \hline
19 & 47991316 & TTAAA/T & upstream & 23\\ \hline
5 & 123647724 & GACA/G & downstream & 22.9\\ \hline
5 & 139749184 & G/A & downstream & 22.9\\ \hline
7 & 27197601 & C/G & downstream & 22.9\\ \hline
8 & 80678427 & T/C & downstream & 22.9\\ \hline
11 & 61513671 & CAGAG/C & downstream & 22.9\\ \hline
11 & 125830697 & T/C & downstream & 22.9\\ \hline
12 & 48396920 & G/T & downstream & 22.9\\ \hline
15 & 100514417 & C/T & downstream & 22.9\\ \hline
16 & 31000726 & TAGAAATGGGACTCTGAGGGCTAAC/T & downstream & 22.9\\ \hline
1 & 41314741 & CA/C & intergenic & 22.9\\ \hline
1 & 92081030 & T/A & intergenic & 22.9\\ \hline
1 & 96499830 & T/A & intergenic & 22.9\\ \hline
2 & 216776870 & T/A & intergenic & 22.9\\ \hline
3 & 6513635 & CGACA/C & intergenic & 22.9\\ \hline
3 & 44148568 & ATGCT/A & intergenic & 22.9\\ \hline
4 & 31809946 & C/CACAC & intergenic & 22.9\\ \hline
5 & 31024929 & TA/T & intergenic & 22.9\\ \hline
6 & 40146909 & ATAAAT/A & intergenic & 22.9\\ \hline
7 & 9863374 & TA/T & intergenic & 22.9\\ \hline
7 & 27252670 & A/G & intergenic & 22.9\\ \hline
10 & 86871665 & TA/T & intergenic & 22.9\\ \hline
11 & 42520473 & T/A & intergenic & 22.9\\ \hline
15 & 95252667 & T/A & intergenic & 22.9\\ \hline
1 & 40155486 & A/G & upstream & 22.9\\ \hline
1 & 48965535 & TAAAC/T & upstream & 22.9\\ \hline
2 & 219722664 & T/C & upstream & 22.9\\ \hline
5 & 117901084 & T/A & upstream & 22.9\\ \hline
7 & 20373528 & TTA/T & upstream & 22.9\\ \hline
7 & 44618599 & T/A & upstream & 22.9\\ \hline
9 & 107694522 & C/T & upstream & 22.9\\ \hline
11 & 63869107 & T/A & upstream & 22.9\\ \hline
1 & 3801973 & T/C & downstream & 22.8\\ \hline
4 & 61659482 & C/CCTT & downstream & 22.8\\ \hline
5 & 123799069 & C/G & downstream & 22.8\\ \hline
5 & 156186841 & G/A & downstream & 22.8\\ \hline
7 & 115147752 & C/A & downstream & 22.8\\ \hline
8 & 28911173 & AGACT/A & downstream & 22.8\\ \hline
9 & 127642032 & CAT/C & downstream & 22.8\\ \hline
12 & 6054748 & T/C & downstream & 22.8\\ \hline
12 & 53806714 & CTG/C & downstream & 22.8\\ \hline
14 & 35903276 & GGACATTTA/G & downstream & 22.8\\ \hline
15 & 92716979 & T/G & downstream & 22.8\\ \hline
2 & 4155192 & T/A & intergenic & 22.8\\ \hline
2 & 143562694 & T/G & intergenic & 22.8\\ \hline
2 & 222173880 & T/A & intergenic & 22.8\\ \hline
4 & 83012112 & T/A & intergenic & 22.8\\ \hline
4 & 83119414 & T/A & intergenic & 22.8\\ \hline
5 & 107795557 & T/A & intergenic & 22.8\\ \hline
5 & 116888166 & T/A & intergenic & 22.8\\ \hline
5 & 166025268 & C/T & intergenic & 22.8\\ \hline
7 & 25314445 & T/A & intergenic & 22.8\\ \hline
10 & 59358896 & T/A & intergenic & 22.8\\ \hline
9 & 75829027 & T/A & intergenic & 22.8\\ \hline
9 & 116544406 & T/G & intergenic & 22.8\\ \hline
11 & 58232952 & T/A & intergenic & 22.8\\ \hline
%\end{tabular}
\end{longtable}